\documentclass[aps,prl,twocolumn,superscriptaddress,showpacs]{revtex4}
\usepackage{graphicx}
\begin{document}

%\preprint{}

\title{Strong parity mixing in the FFLO superconductivity in 
systems with coexisting spin and charge fluctuations}
% with long range off-site repulsions}



\author{Hirohito Aizawa}
%\email[Electronic address: ]{aizawa@vivace.e-one.uec.ac.jp}
%\thanks{}
\affiliation{
 Department of Applied Physics and Chemistry, The University of
 Electro-Communications, Chofu, Tokyo 182-8585, Japan}

\author{Kazuhiko Kuroki}
%\thanks{}
\affiliation{
 Department of Applied Physics and Chemistry, The University of
 Electro-Communications, Chofu, Tokyo 182-8585, Japan}

\author{Takehito Yokoyama}
%\thanks{}
\affiliation{
 Department of Applied Physics, Nagoya University,
 Nagoya 464-8603, Japan}

\author{Yukio Tanaka}
%\thanks{}
\affiliation{
 Department of Applied Physics, Nagoya University,
 Nagoya 464-8603, Japan}


\date{\today}

\begin{abstract}
We study the Fulde-Ferrell-Larkin-Ovchinnikov (FFLO) state of 
spin fluctuation mediated pairing, and focus on the effect of 
coexisting charge fluctuations.
We find that 
 (i) consecutive transitions from 
singlet pairing to FFLO and further 
to $S_z=1$ triplet pairing  can generally take place 
upon increasing the magnetic field 
when strong charge fluctuations coexist with spin fluctuations, 
and
 (ii) the enhancement of the charge fluctuations 
lead to a significant increase of the parity mixing in the FFLO state, 
where the triplet/singlet component ratio in the gap function 
can be close to unity.
 We propose that such consecutive pairing state transition and 
strong parity mixing in the FFLO state may take place in 
a quasi-one-dimensional organic superconductor (TMTSF)$_2X$.
\end{abstract}

%\keywords{}

\maketitle

%%%%%%%%%%%%%%%%%%%%%%%%%%%%%%%%%%%%%%%%%%%%%%%%%%%%%%%%%%%%%%%%%%%%%%%%%%
%% body of paper here - Use proper section commands
%% References should be done using the \cite, \ref, and \label commands
%%
%\section{Introduction\label{Introduction}}
%%
 The Fulde-Ferrell-Larkin-Ovchinnikov (FFLO) state, 
 in which the Cooper pairs 
 formed as $( k+Q_{c}\uparrow, -k+Q_{c}\downarrow)$ 
 have a finite center of mass momentum, 
 is one of the most fascinating superconducting states. 
 \cite{Fulde-Ferrell, Larkin-Ovchinnikov} 
 One of the interesting aspects of the FFLO state is the parity mixing, 
 i.e.,  even and odd parity pairings can be mixed. 
 Phenomenological studies have shown that the mixing of 
 the singlet and triplet pairings stabilizes the FFLO state. 
 \cite{Matsuo-Shimahara-Nagai,Shimahara2} 
 Recent microscopic studies 
 have shown that the $S_{z}=0$ triplet pairing is mixed with singlet 
 pairing in the FFLO state 
 of the Hubbard model on the two-leg ladder-type lattice, 
 \cite{Roux-White-Capponi-Poilblanc} 
 and also on the square lattice, 
 where $d$-wave superconductivity is mediated by spin fluctuations. 
 \cite{Yanase, Yokoyama-Onari-Tanaka-FFLO} 
In ref.\cite{Yanase} it has been pointed out 
that the parity mixing stabilizes the FFLO state, 
even in the vicinity of the quantum critical point where the 
quasi-particle lifetime becomes short due to the scattering by 
spin fluctuations.

 Recent experimental indications of possible occurrence of the FFLO state 
 in CeCoIn$_{5}$, 
 \cite{ Radovan-Fortune-et-al} 
 a quasi-two-dimensional (Q2D) organic materials such as   
 $\lambda$-(BETS)$_{2}$FeCl$_{4}$ 
 (BETS=bisethylenedithio-tetraselenafulvalene) 
 \cite{Uji-Terashima-et-al}  and  
 $\kappa$-(BEDT-TTF)$_{2}$Cu(NCS)$_{2}$ 
 (BEDT-TTF=bisethylenedithio-tetrathiafulvalene), 
 \cite{Singleton-Symington} % \cite{Singleton-Symington, Lortz-Wang-et-al}
 and also in a quasi-one-dimensional(Q1D) one 
 (TMTSF)$_{2}$ClO$_{4}$ 
 (TMTSF=tetramethyl-tetraselenafulvalene) 
 \cite{Shinagawa-Kurosaki-et-al, Yonezawa-Kusaba-et-al} 
 have stimulated extensive studies in this field. 
 For (TMTSF)$_{2}$ClO$_{4}$ in particular, 
 the possibility of the spin triplet pairing has previously 
 been suggested experimentally for (TMTSF)$_{2}X$ 
 ($X$=PF$_{6}$, 
 \cite{ Lee-Naughton-et-al, Lee-Brown-et-al} 
 ClO$_{4}$ \cite{Oh-Naughton, Shinagawa-Wu}), 
% \cite{ Lee-Naughton-et-al-Lee-Chaikin-et-al, 
% Lee-Brown-et-al-Lee-Chow-et-al, Lee-review} 
 but a more recent NMR experiment  
by Shinagawa {\it et al.} \cite{Shinagawa-Kurosaki-et-al}
 has revealed that when the magnetic 
field is low, the pairing occurs in the spin-singlet channel, 
while when the magnetic field is high, the pairing state is either
an FFLO state or a spin-triplet state.
Yonezawa {\it et al.} \cite{Yonezawa-Kusaba-et-al}
have found that the onset $T_c$ exhibits a peculiar magnetic 
field direction dependence at high fields, which may be 
related to the occurrence of the FFLO state, where the 
direction of the total momentum of the Cooper pairs can play 
an important role. 
%Moreover, Yonezawa {\it et. al} %\cite{Yonezawa-Kusaba-et-al}
% have pointed out, by comparing samples with different quality, 
% that there may be two kinds of high field pairing states.
 However, the microscopic origin of the transition between 
 high field pairing states 
 like the FFLO or the triplet state 
%, like from the FFLO to the triplet state, 
 remains unexplored.

Theoretically, various studies have investigated the possibility 
of triplet pairing 
 \cite{Lebed-FIDC, Fuseya-Suzumura, Nickel-Duprat, 
% Shimahara-FIST, Vaccarella-Melo, Fuseya-Onishi-Kohno-et-al, 
% Belmechri-Abramovici-et-al, Belmechri-Abramovici-Heritier}
 Shimahara-FIST, Vaccarella-Melo, Belmechri-Abramovici-et-al} 
and the FFLO state
 \cite{Dupuis-Montambaux-Melo, Miyazaki-Kishigi-Hasegawa}. 
 In particular, 
 three of the present authors 
 have previously shown that the triplet $f$-wave pairing can compete 
 with the singlet $d$-wave pairing 
 in the Q1D system 
 because of the disconnectivity of the Fermi surface 
 when $2k_{F}$ spin and $2k_{F}$ charge fluctuations coexist. 
 \cite{Kuroki-Arita-Aoki, Tanaka-Kuroki-and-Kuroki-Tanaka, Kuroki-review}
% The coexistence of the charge fluctuations and the spin fluctuations is 
 $2k_{F}$ spin+$2k_{F}$ charge fluctuations 
 supported from the fact that 
 diffuse X-ray scattering experiments observe the coexistence of 
 $2k_{F}$ charge density wave(CDW) and the 
 $2k_{F}$ spin density wave(SDW) in the 
 vicinity of the superconducting phase 
 in (TMTSF)$_{2}$PF$_{6}$. 
 \cite{Pouget-Ravy, Kagoshima-Saso-et-al} 
% The origin of the competition between the spin triplet $f$-wave 
% and the spin singlet $d$-wave is a disconnectivity of 
% a Fermi surface in the Q1D system, since one of the nodes 
% in the spin triplet gap is absent on the Fermi surface.  
%
% Recent study for the temperature dependence of NMR $1/T_{1}$ 
% Recent theoretical study for NMR $1/T_{1}$ 
% on the model for (TMTSF)$_{2}X$ 
% on (TMTSF)$_{2}X$ 
% \cite{Takigawa-Ichioka-et-al} 
% has shown that 
% the $d$-wave or $f$-wave %with gap nodes on the Fermi surface 
% almost agrees with the experiments 
% NMR 1/T1 experiments 
% \cite{ Lee-Brown-et-al, Shinagawa-Kurosaki-et-al}
% \cite{Takigawa-Yasuoka-et-al, Lee-Brown-et-al-Lee-Chow-et-al, 
% Shinagawa-Kurosaki-et-al}% NMR 1/T1 experiments 
% rather than $p$-wave,  
% which indicate nodes of the gap on the Fermi surface 
% although a thermal conductivity suggests the nodeless gap. 
% \cite{Belin-Behnia} 
% 
% and it seems that the difference between the $d$-wave and the $f$-wave 
% is unobservable from the NMR $1/T_{1}$ results. 
% \cite{Takigawa-Ichioka-et-al} 
%
%
 Moreover, we have recently found that this kind of triplet pairing due
 to %the coexistence of 
 $2k_{F}$ spin+$2k_{F}$ charge fluctuations is strongly enhanced 
 by the magnetic field.
%, which supports the singlet-triplet transition 
% in (TMTSF)$_{2}X$ under the field. 
% under the field. 
 \cite{Aizawa-Kuroki-Tanaka} 
 Then a naive question arises along this line: 
 what happens if magnetic field is applied to a system where 
 spin singlet pairing dominates at zero field but triplet pairing 
 is closely competing ? 
 If the FFLO state emerges, what is its nature ? 

% \cite{Dupuis-Montambaux-Melo, Dupuis-Montambaux, 
% Miyazaki-Kishigi-Hasegawa}. 
% and 1D \cite{Suzumura-Ishino, Machida-Nakanishi}. 

 Given this background, in the present Letter, 
 we study the FFLO state of spin fluctuation 
 mediated superconductivity in low dimensional systems, 
 and focus on the  effect of the charge fluctuations. 
 We find that 
 (i) consecutive transitions from 
 singlet pairing to FFLO and further 
 to $S_z=1$ triplet pairing  can generally take place 
 upon increasing the magnetic field 
 in the vicinity of the SDW+CDW coexisting phase, 
 and 
 (ii)
 the enhancement of the charge fluctuations 
 leads to a significant increase of the parity mixing in the FFLO state, 
 where the triplet/singlet component ratio in the gap function 
 can be close to unity. 
 Based on a calculation on a model for (TMTSF)$_2X$, 
 we propose that such consecutive pairing state transitions 
 and the strong parity mixing in the FFLO state may actually 
 be taking place in this material.

\begin{figure}[!htb]
  \includegraphics[width=7.0cm]{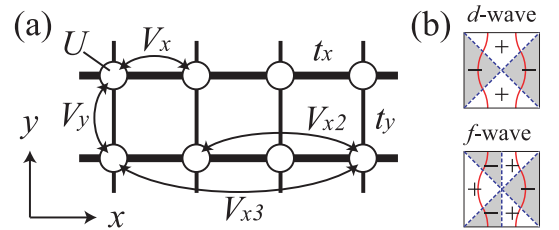}
  \caption{
  (a) The model adopted in this study. 
  (b) The schematic figure of the gap for $d$-wave(upper) 
  and $f$-wave(lower), 
  where the nodes of gap (blue dashed lines), 
  the disconnected Fermi surface in Q1D lattice (red solid curves). 
%  the Fermi surface for square (red solid curve) and 
%  Q1D (green long dashed curves) lattice.
  }
  \label{model}
 \end{figure}

The anisotropic extended Hubbard model [Fig.\ref{model} (a)] 
that takes into account the Zeeman effect is given by 
\begin{eqnarray}
 H = 
  \sum_{i,j,\sigma} t_{ij\sigma} c_{i\sigma}^\dagger c_{j\sigma}
  +\sum_{i} U n_{i\uparrow}  n_{i\downarrow} 
%  +\sum_{i,j,\sigma,\sigma'} V_{ij} n_{i\sigma} n_{j\sigma'}. 
  +\sum_{i,j} V_{ij} n_{i} n_{j}. 
 \label{hamiltonian}
\end{eqnarray}
Here $t_{ij\sigma}=t_{ij}+h_{z}{\rm sgn}(\sigma)\delta_{ij}$, where 
the hopping $t_{ij}$ is considered only for intrachain ($t_x$) and the 
interchain ($t_y$) nearest neighbors. $t_{x}=1.0$ is taken as  
the energy unit. 
 $U$ is the on-site repulsion, and $V_{ij}$ are the off-site repulsions: 
$V_x$, $V_{x2}$, $V_{x3}$ are nearest, next nearest and 3rd nearest 
 neighbor interaction within the chains, 
 and $V_y$ is the interchain interaction.
 We ignore the orbital effect, 
 assuming that the magnetic field is applied parallel to the conductive 
 $x$-$y$ plane, 
 thus we assume a sufficiently large Maki parameter. 

 The bare susceptibilities, bubble-type and ladder-type, are written as 
\begin{eqnarray}
 \chi_{0}^{\sigma \sigma}(k)
  &=&\frac{-1}{N}\sum_{q}
  \frac{f(\xi_{\sigma}(k+q))-f(\xi_{\sigma}(q))}
       {\xi_{\sigma}(k+q)-\xi_{\sigma}(q)},
  \label{a-chi0-para}
\\
 \chi_{0}^{+-}(k)
  &=&\frac{-1}{N}\sum_{q}
  \frac{f(\xi_{\sigma}(k+q))-f(\xi_{\bar{\sigma}}(q))}
       {\xi_{\sigma}(k+q)-\xi_{\bar{\sigma}}(q)},
  \label{a-chi0-pm}
\end{eqnarray}
 where $\xi_{\sigma}(k)$ is the band dispersion that takes into account  
 the Zeeman effect measured from the chemical potential $\mu$, 
 and $f(\xi)$ is the Fermi distribution function.

 Within RPA that takes into account the magnetic field parallel 
 to the spin quantization axis $\hat{z}$, 
 \cite{Aizawa-Kuroki-Tanaka} 
 the longitudinal spin and charge susceptibilities are given as 
 $\chi_{sp}^{zz}=\frac{1}{2}
(\chi^{\uparrow \uparrow}+\chi^{\downarrow \downarrow}
-\chi^{\uparrow \downarrow}-\chi^{\downarrow \uparrow})$ and
 $\chi_{ch}=\frac{1}{2}
(\chi^{\uparrow \uparrow}+\chi^{\downarrow \downarrow}
+\chi^{\uparrow \downarrow}+\chi^{\downarrow \uparrow})$, where 
\begin{eqnarray}
 \chi^{\sigma \sigma}(k) &=&
  \left[ 1+\chi_{0}^{{\bar \sigma} {\bar \sigma}}(k) V(k)\right] 
  \chi_{0}^{\sigma \sigma}(k)/A(k), 
\label{a-chi-para}\\
 \chi^{\sigma {\bar \sigma}}(k) &=&
  -\chi_{0}^{\sigma \sigma}(k)
  \left[U+V(k)\right]\chi_{0}^{{\bar \sigma} {\bar \sigma}}(k)/A(k), 
\label{a-chi-anti-para}\\
 A(k) &=& \left[ 1+\chi_{0}^{\sigma \sigma}(k) V(k) \right]
    \left[ 1+\chi_{0}^{{\bar \sigma} {\bar \sigma}}(k) V(k)\right]
 \nonumber \\ & &
    -\left[U+V(k)\right]^{2}
    \chi_{0}^{\sigma \sigma}(k) 
    \chi_{0}^{{\bar \sigma} {\bar \sigma}}(k). 
\end{eqnarray}
 The transverse spin susceptibility, 
 in which we ignore the off-site repulsions for simplicity, 
 is given as
\begin{eqnarray}
 \chi_{sp}^{+-}(k)=\frac{\chi_{0}^{+-}(k)}{1-U\chi_{0}^{+-}(k)}. 
\label{a-chipm}
\end{eqnarray}

 The pairing interactions from the bubble and ladder diagrams
 are given as 
\begin{eqnarray}
 V^{\sigma \bar{\sigma}}_{bub}(k)
 &=& 
 U+V(k)+\frac{U^{2}}{2}\chi_{sp}^{zz}(k)
 \nonumber \\ & &
 -\frac{\left[ U+2V(k) \right]^{2}}{2}\chi_{ch}(k),
 \label{a-Vs-bub}\\
 V^{\sigma \bar{\sigma}}_{lad}(k)
 &=& 
 U^{2}\chi_{sp}^{+-}(k),
 \label{a-Vs-lad}\\
 V^{\sigma \sigma}_{bub}(k) 
 &=& 
 V(k)-2\left[ U+V(k) \right]V(k)\chi^{{\sigma} \bar{\sigma}}(k)
 \nonumber \\ & &
     -V(k)^{2}\chi^{{\sigma} {\sigma}}(k)
     -\left[ U+V(k) \right]^{2}\chi^{\bar{\sigma} \bar{\sigma}}(k),
 \label{a-Vt-para-bub}\\
 V^{\sigma \sigma}_{lad}(k) 
 &=& 0.
 \label{a-Vt-para-lad}
\end{eqnarray}
% for the opposite spins ($\sigma \bar{\sigma}$) which 
% is the singlet and/or triplet with $S_{z}=0$ pairing
% and the equal spins ($\sigma \sigma$) which 
% is the triplet with $S_{z}=\pm 1$ pairing, respectively. 

 The linearized gap equation for Cooper pairs with
 the total momentum $2Q_{c}$ 
 ($Q_{c}$ represents the center of mass momentum) 
 is given by 
\begin{eqnarray}
 \lambda^{\sigma \sigma'}_{Q_{c}} \phi^{\sigma \sigma'}(k)
  = \frac{1}{N}\sum_{q}
  [V^{\sigma \sigma'}_{bub}(k-q)+V^{\sigma \sigma'}_{lad}(k+q)]
  \nonumber \\  
 \times
  \frac{ f(\xi_{\sigma}(q_{+})) 
        -f(-\xi_{\sigma'}(-q_{-}))}
       {\xi_{\sigma}(q_{+})+\xi_{\sigma'}(-q_{-})}
  \phi^{\sigma \sigma'}(q), 
 \label{gap-eq}
\end{eqnarray}
 where $q_{\pm}=q \pm Q_{c}$, 
 $\phi^{\sigma \sigma'}(k)$ is the gap function 
 and $\lambda^{\sigma \sigma'}_{Q_{c}}$ is the eigenvalue of this 
 linearized gap equation.
 The center of mass momentum ${\bf Q}_{c}$ which gives the maximum value 
 of $\lambda^{\sigma \bar{\sigma}}_{Q_{c}}$ lies along the $x$-direction 
 because of the nesting of the Fermi surface 
 \cite{Shimahara-FFLO_direction}
 and $\lambda^{\sigma \sigma}_{Q_{c}}$ takes its maximum  
 at ${\bf Q}_{c}=(0,0)$ 
 because the electrons do not scatter between the different directional 
 spins in this pairing channel.

%, note that we confirm 
% the $Q_{c}$-dependence of the eigen value $\lambda^{\sigma \sigma'}$. 
% Satisfying the condition of 
% $\lambda^{\sigma \sigma'}_{Q_{c}\ne 0} > 
%  \lambda^{\sigma \sigma'}_{Q_{c}= 0}$, 
% the FFLO state with $Q_{c}$ 
% is dominant over the usual pairing state. 
% is dominant over the usual opposite spin pairing state. 
% with spin ($\sigma \sigma'$). 
 We define the singlet and the $S_{z}=0$ triplet component 
 of the gap function in the opposite spin pairing channel as 
\begin{eqnarray}
 \phi_{{\rm SS}}(k)&=&
  \left[\phi^{\uparrow \downarrow}(k)
   -\phi^{\downarrow \uparrow}(k)\right]/2, 
\nonumber \\
% \phi_{{\rm ST}^{\uparrow \downarrow}}(k)&=&
 \phi_{{\rm ST}^{0}}(k)&=&
  \left[\phi^{\uparrow \downarrow}(k)
   +\phi^{\downarrow \uparrow}(k)\right]/2.
 \label{eq-phis-phit}
\end{eqnarray} 
 In our calculation, 
 the spin singlet and the spin triplet component of the gap function 
 in the FFLO state is essentially 
 $d$-wave and $f$-wave 
 as schematically shown in Fig \ref{model} (b), 
% $d_{x^{2}-y^{2}}$-wave and $f_{x^{3}-xy^{2}}$-wave 
% as schematically shown in Fig, \ref{model} (b), 
% thus we omit the former as $d$-wave and the latter as $f$-wave   
 so we write the singlet ($S_{z}=0$ triplet) component of the FFLO gap 
  $\phi_{{\rm SS}}$($\phi_{{\rm ST}^{0}}$) 
%  $\phi_{{\rm SS}}$($\phi_{{\rm ST}^{\uparrow \downarrow}}$) 
 in Eq. (\ref{eq-phis-phit}) as 
 $\phi_{{\rm SS}d}$ ($\phi_{{\rm ST}f^{0}}$), 
% $\phi_{{\rm SS}d}$ ($\phi_{{\rm ST}f^{\uparrow \downarrow}}$), 
 where SS$d$(ST$f^{0}$) stands for 
% where SS$d$(ST$f^{\uparrow \downarrow}$) stands for 
 spin singlet $d$-wave (spin triplet $f$-wave with $S_{z}=0$) pairing. 
 The eigenvalue of each pairing state is determined as follows.
$\lambda^{\sigma \bar{\sigma}}_{Q_{c}}$ 
 with ${\bf Q}_{c}=(0,0)$ gives 
the eigenvalue of the singlet $d$-wave pairing $\lambda_{{\rm SS}d}$ 
($S_z=0$ triplet $f$-wave pairing 
 $\lambda_{{\rm ST}f^{0}}$)
% $\lambda_{{\rm ST}f^{\uparrow \downarrow}}$)
 $\phi_{{\rm ST}f^{0}}=0$ ($\phi_{{\rm SS}d}=0$),
% $\phi_{{\rm ST}f^{\uparrow \downarrow}}=0$ ($\phi_{{\rm SS}d}=0$),
while $\lambda^{\sigma \bar{\sigma}}_{Q_{c}}$ 
 with ${\bf Q}_{c}\ne (0,0)$ gives $\lambda_{{\rm FFLO}}$.
$\lambda^{\sigma \sigma}_{Q_{c}}$  with ${\bf Q}_{c}=(0,0)$ gives  
the eigenvalue for the spin triplet $f$-wave 
pairing with $S_{z}=+1$ ($S_{z}=-1$)
 $\lambda_{{\rm ST}f^{+1}}$ ($\lambda_{{\rm ST}f^{-1}}$).
% $\lambda_{{\rm ST}f^{\uparrow \uparrow}}$
% ($\lambda_{{\rm ST}f^{\downarrow \downarrow}}$).
%
%
%% Although RPA may be considered as quantitatively insufficient 
%% for discussing the absolute value of $T_{c}$ and 
%% the value of order parameter, 
%% we expect this approach to be valid for studying 
%% the effect of the dimensionality and the charge fluctuations on 
%% the {\it competition} between different pairing symmetries such as  
%% the singlet, triplet and FFLO state. 
% in not so large magnetic field regime.

%%%%%%%%%%%%%%%%%%%%%%%%%%%%%%%%%%%%%%%%%%%%%%%%%%%%%%%%%%%%%%%%%%%%%%%%%%
%%
%\section{Results - nearly half-filling - \label{results-n11}}

 First, to make the argument general, 
 we concentrate on a simple model with only the on-site $U=1.5$ 
 and the nearest neighbor repulsion $V_x$ in the $x$-direction.  
% @
When $V_x$ is large, $2k_F$ charge fluctuations tends to develop for  
band fillings close to half filling, so 
we take the band filling $n=1.1$,  
where $n=$number of electrons/number of sites.   
%@
Here we fix the value of $t_y$ at 0.5, 
but this value does not have a specific meaning, 
and qualitatively (although not quantitatively)  
similar results can be obtained for other 
values of $t_y$. % is this OK? 
The temperature is fixed at $T=0.01$ here.
 System size is taken as 2048$\times$64 sites here. 
 
\begin{figure}[!htb]
  \includegraphics[width=7.5cm]{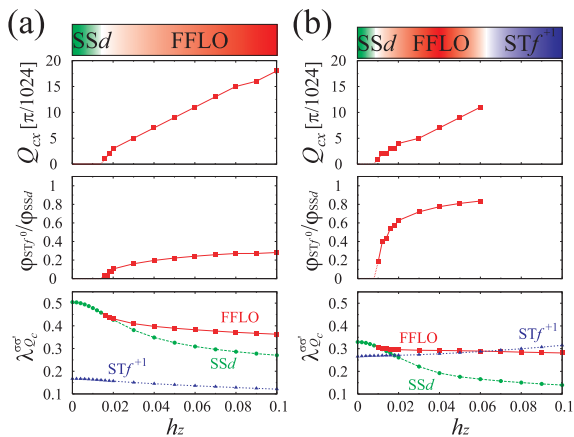}
  \caption{(Color online) 
  The $h_{z}$-dependence of $Q_{cx}$(upper panels), 
  the $S_{z}=0$ triplet/singlet ratio in the FFLO 
  $\phi_{{\rm ST}f^{0}}/\phi_{{\rm SS}d}$(middle) 
  and the gap equation eigenvalues (lower) for 
  (a) $V_{x}=0$ and (b) $V_{x}=0.65$. 
  Other parameters are $U=1.5$, $n=1.1$, $t_y=0.5$, and $T=0.01$.
  }
  \label{hz-dep-n11}
 \end{figure}

% @
In Fig.\ref{hz-dep-n11}, 
 we show the magnetic field dependence of 
 $Q_{cx}$ of the FFLO state, 
 the parity mixing rate 
 $\phi_{{\rm ST}f^{0}}/\phi_{{\rm SS}d}$ 
 and the eigenvalues of the gap equation 
 for (a) $V_x=0$ and (b) $V_x=0.65$. 
 Note that we denote 
 the ratio between the maximum value of the $S_z=0$ triplet component 
 and that of the singlet component of the gap function 
 in the FFLO state as 
 ``$\phi_{{\rm ST}f^{0}}/\phi_{{\rm SS}d}$'' 
 hereafter. 
%@
 The dominating pairing state changes from singlet $d$-wave 
 to FFLO upon increasing the magnetic field 
 for both $V_{x}=0$ and $V_{x}=0.65$, 
 but for sufficiently large field, 
 FFLO further gives way to the triplet $f$-wave state with $S_z=1$ 
for $V_{x}=0.65$, i.e., when the charge fluctuations are present. 
%
%@@ The reason why $S_{z}=1$ triplet increases.
%@@@@
 The reason why $S_{z}=1$ triplet $f$-wave dominates at high fields can be 
 explained as follows. 
 The presence of the charge fluctuations suppresses 
 the spin singlet pairing interaction and enhance the triplet one. 
 \cite{Kuroki-Arita-Aoki, Tanaka-Kuroki-and-Kuroki-Tanaka,  
 Kuroki-review, Fuseya-Suzumura, Nickel-Duprat} 
 Secondly, 
 an $S_z=1$ triplet pairing state induced by the coexistence of 
 spin and charge fluctuations is strongly enhanced by the magnetic field 
 applied parallel to the spin quantization axis $\hat{z}$. 
 \cite{Aizawa-Kuroki-Tanaka}
%@@@@
%@@@
%The reason why $S_{z}=1$ triplet $f$-wave dominates at high fields can be 
%traced back in previous studies.
%First of all, the presence of the charge fluctuations suppresses the 
%spin singlet pairing interaction and enhance the triplet one, 
%making the triplet pairing channel more 
%competitive.\cite{Kuroki-Arita-Aoki, Tanaka-Kuroki-and-Kuroki-Tanaka,  
% Kuroki-review, Fuseya-Suzumura, Nickel-Duprat} Secondly, 
%an $S_z=1$ triplet pairing state induced by the coexistence of 
%spin and charge fluctuations is strongly enhanced by the magnetic field 
%the applied in the $z$ direction.\cite{Aizawa-Kuroki-Tanaka}
%@@@
% The bubble diagrams at the nesting vector 
% $\textbf{\textit{Q}}_{2k_{F}}$, 
% $\chi_{0}^{\uparrow \uparrow}
%\left(\textbf{\textit{Q}}_{2k_{F}}\right)$, 
% enter as the unpaired form in the $S_{z}=+1$ triplet pairing channel 
% because they are connected by off-site interactions 
% $V\left( \textbf{\textit{Q}}_{2k_{F}}\right)$ 
% when the $2k_{F}$ spin+$2k_{F}$ charge fluctuations, 
% although the bubble diagrams in the $S_{z}=+1$ triplet channel always 
% enter as paired units consisting of 
% $\chi_{0}^{\uparrow \uparrow}
% \left(\textbf{\textit{Q}}_{2k_{F}}\right)
% \chi_{0}^{\downarrow \downarrow}
% \left(\textbf{\textit{Q}}_{2k_{F}}\right)$
% because of the Pauli principle 
% when off-site repulsions are absent, thus, 
% solely spin fluctuations is present. 
%@@
%@@ The reason why $S_{z}=1$ triplet increases.
%
In fact, such a possibility of transition from singlet pairing 
to FFLO, and further to triplet pairing upon increasing the 
magnetic field has been phenomenologically 
proposed by Shimahara.\cite{Shimahara-FIST}
% The effect of the dimensionality and the charge fluctuations 
% is controlled 
% by using the parameters as the transfer in $y$-direction $t_{y}$ and 
% the first nearest neighbor (NN) off-site repulsions $V_{x}$. 

\begin{figure}[!htb]
  \includegraphics[width=6.0cm]{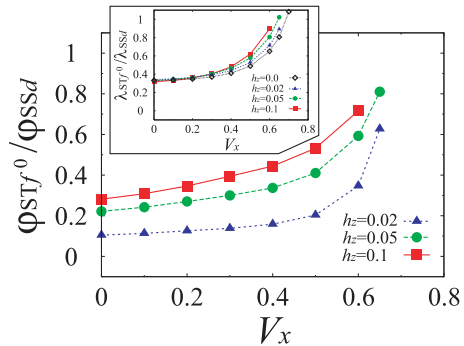}
  \caption{(Color online) 
%  The $V_{x}$-dependence of  (a) the ratio
  The $V_{x}$-dependence of the ratio
  $\phi_{{\rm ST}f^{0}}/\phi_{{\rm SS}d}$ and 
%  (b) the ratio of the eigenvalues between the $S_{z}=0$ triplet and 
%  the singlet. 
  the ratio of the eigenvalues between the $S_{z}=0$ triplet and 
  the singlet on inset figure. 
  Other parameters are the same with Fig.\ref{hz-dep-n11}. 
  }
  \label{Vx-dep}
 \end{figure}

 Let us now look into the nature of the FFLO state.
 We show the $V_{x}$ dependence of the ratio 
 $\phi_{{\rm ST}f^{0}}/\phi_{{\rm SS}d}$ 
% in Fig.\ref{Vx-dep}(a). 
 in Fig.\ref{Vx-dep}. 
 We see that the mixing ratio increases as $V_x$, namely, 
the charge fluctuation increases.
 The strong mixing of the triplet pairing component 
 may be expected from the fact that 
 the presence of charge fluctuations makes triplet pairing 
 more competitive against singlet pairing, 
 \cite{Kuroki-Arita-Aoki, Tanaka-Kuroki-and-Kuroki-Tanaka,  
 Kuroki-review, Fuseya-Suzumura, Nickel-Duprat, Aizawa-Kuroki-Tanaka} 
%@
%This can be seen clearly in Fig.\ref{Vx-dep}(b), where 
This can be seen clearly in inset figure on Fig.\ref{Vx-dep}, where 
we plot the eigenvalue of the singlet $d$-wave and 
$S_z=0$ triplet $f$-wave pairing using the 
formalism adopted in ref.\cite{Aizawa-Kuroki-Tanaka}, where 
mixing between the odd and even parity pairings is prohibited.
%
% Next we show the $V_{x}$ dependence of the eigenvalue $\lambda$ 
% for the each pairing channels in Fig.\ref{Vx-dep}(b) 
% for $t_y=0.5$ and $h_{z}=0.05$.  
% Increasing the $V_{x}$ 
% suppresses both the singlet $d$-wave and the FFLO state, 
% however, the suppression in the FFLO state 
% is slightly smaller than that in the singlet $d$-wave. 
% Namely, the FFLO state is relatively favored rather than 
% the singlet $d$-wave pairing state with increasing the $V_{x}$. 
% In large $V_{x}$ regime as for the $S_{z}=1$ triplet is most dominant, 
% large triplet/singlet pairty mixing ratio leads to
% the $S_{z}=0$ triplet $f$-wave from the singlet $d$-wave 
% in the opposite spin pairing channel with the $Qcx=0$. 
% Accordingly, the FFLO state with large $S_{z}=0$ triplet component 
% arises.  
% However, the $S_{z}=0$ triplet dominant FFLO state is always smaller 
% than the $S_{z}=1$ triplet pairing. 

%%%%%%%%%%%%%%%%%%%%%%%%%%%%%%%%%%%%%%%%%%%%%%%%%%%%%%%%%%%%%%%%%%%%%%%%%%%%%%
%%
%\section{Results - (TMTSF)$_{2}$X - \label{results-TMTSF}}

 We now move on to a realistic model for (TMTSF)$_{2}X$.
 We introduce not only $U$ and $V_x$, but also other 
 distant off-site repulsions  $V_{x2}$, $V_{x3}$ and $V_{y}$,  
 since the $2k_{F}$ charge fluctuations, 
 which becomes competitive against $2k_F$ spin fluctuations  
when the condition of $V_{x2}+V_{y} \simeq U/2$ 
 is satisfied, are important in this material. 
 \cite{Kuroki-Arita-Aoki, Tanaka-Kuroki-and-Kuroki-Tanaka, 
 Kuroki-review, Aizawa-Kuroki-Tanaka} 
 Here, we fix the repulsions as $U=1.7$, $V_{x}=0.9$, $V_{x2}=0.45$ 
 and $V_{x3}=0.1$, and  vary $V_{y}$. 
 Other parameters are taken as $t_{y}=0.2$, $T=0.012$, $n=1.5$ (3/4 filling), 
 and the system size is taken as 1024$\times$64.

\begin{figure}[!htb]
  \includegraphics[width=7.5cm]{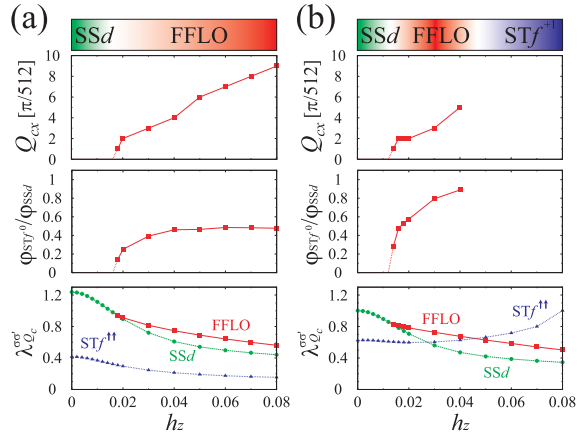}
  \caption{(Color online) 
  The $h_{z}$-dependence of $Q_{cx}$(upper), 
  the triplet/singlet ratio 
  $\phi_{{\rm ST}f^{0}}/\phi_{{\rm SS}d}$(center) 
  and the each eigenvalues (lower) in 
  (a) the absence and (b) the presence of the off-site repulsions. 
  }
  \label{hz-dep-n15}
 \end{figure}

 In Fig. \ref{hz-dep-n15},  
 the center of mass momentum $Q_{cx}$, 
 the parity mixing ratio
 $\phi_{{\rm ST}f^{0}}/\phi_{{\rm SS}d}$ 
 and the eigenvalues 
 are plotted as functions of $h_{z}$ in 
 the (a) absence ($V_{x}$, $V_{x2}$, $V_{x3}$ and $V_{y}=0$) 
 or (b) presence of the off-site repulsions, where  
 for the latter case we set $V_{y}=0.35$, 
 for which $2k_{F}$ charge fluctuations are smaller than $2k_{F}$ spin 
 fluctuations. 
 As in the previous case for the simple model, 
 FFLO dominates over singlet $d$-wave pairing upon increasing the 
 magnetic field, 
 and the FFLO state further gives way to the  $S_{z}=1$ triplet $f$-wave
 state in the presence of the off-site repulsions. 
 As shown in the middle panels in Fig.\ref{hz-dep-n15}, 
 a strong mixing of singlet and $S_{z}=0$ triplet pairing components 
 takes place in the FFLO state, 
 especially when the charge fluctuations are strong.

\begin{figure}[!htb]
  \includegraphics[width=8.0cm]{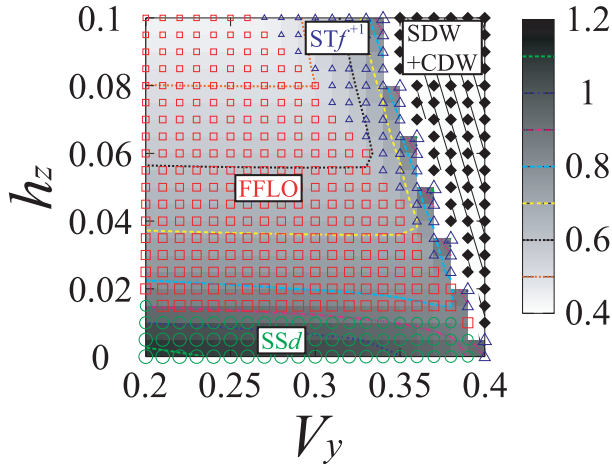}
  \caption{(Color online) 
  A pairing phase diagram in $V_{y}$-$h_{z}$ space, 
  where  the green circles represent the singlet $d$-wave, 
  the red squares the FFLO, 
  the blue triangles the $S_{z}=1$ triplet $f$-wave and 
  the black filled diamonds SDW+CDW. 
  In the shaded area, the eigenvalue is represented by the contours and
  the hatched line area is the SDW+CDW instability. 
  }
  \label{V-hz-pairing-phase}
 \end{figure}

 Finally, we show in Fig. \ref{V-hz-pairing-phase} 
 a phase diagram for the pairing competition
 in the off-site repulsion $V_{y}$ versus magnetic field $h_{z}$ space 
 obtained by comparing the eigenvalues of the gap equation for 
 each pairing channel. 
 The size of the symbols denotes the magnitude of 
 the eigenvalues,  and ``SDW+CDW'' means that both spin 
 and charge susceptibilities have divergently large values.
 The phase diagram shows that for large enough $V_y$, 
 the singlet$\rightarrow$FFLO(with strong 
 parity mixing)$\rightarrow$triplet transition 
 takes place upon increasing the magnetic field.  

%%%%%%%%%%%%%%%%%%%%%%%%%%%%%%%%%%%%%%%%%%%%%%%%%%%%%%%%%%%%%%%%%%%%%%%%%%%%%%
%%
%\section{conclusion\label{conclusion}}

 To conclude, we find that 
 (i) consecutive transitions from singlet pairing to FFLO 
 and further to $S_z=1$ triplet pairing can generally take place 
 upon increasing the magnetic field 
 in the vicinity of the SDW+CDW coexisting phase, and
 (ii) the enhancement of the charge fluctuations leads to 
 a significant increase of the parity mixing in the FFLO state, 
 where the triplet/singlet component ratio in the 
 gap function can be close to unity.

 We raise (TMTSF)$_2X$ as a candidate material 
 for such consecutive pairing state transition and 
 strong parity mixing in the FFLO state to take place.  
%@@@
In fact, as mentioned in the introductory part, the experiments 
 suggest the presence of low field and high field pairing states,
 where the former occurs in the spin-singlet 
channel. \cite{Shinagawa-Kurosaki-et-al} 
 As for the high magnetic field pairing state, Yonezawa {\it et al} 
have shown that for a magnetic field parallel to the $a$ axis, 
only the clean sample exhibits an upturn of the $T_c$ curve 
in the high magnetic field regime above 4T, which suggests the  
presence of a pairing state sensitive to the impurity content. 
\cite{Yonezawa-Kusaba-et-al}
Between 4T and the Pauli limit of around 2.5T, there seems to be 
a different high field pairing state, 
in which superconductivity is stable against the 
impurities, but is very sensitive to the tilt of the magnetic 
field out of the $a$-$b$ plane. The bottom line of 
 these experiments is that 
there may be three kinds of pairing states, i.e., one low field state, 
and two high field states. The correspondence between these 
experimental observations and the present study is not clear 
at the present stage, but the appearance of 
three kinds of pairing states is indeed intriguing.
It would be interesting to further investigate experimentally 
the possibility and the nature of two kinds of 
 high field pairing states.
%@@@
% @@ about correspondence 
%@@
% Finally, we refer correspondences with our results in this letter and 
% the experiments by Yonezawa {\it et. al}. 
% \cite{Yonezawa-Kusaba-et-al} 
% The pairing transition as 
% the singlet $\rightarrow$ 
% FFLO(with strong $S_{z}=0$ triplet mixing) $\rightarrow$ 
% $S_{z}=1$ triplet upon increasing the magnetic field 
% is similar with 
% the experimental measurement of 
% three different superconducting states 
% upon the magnetic field as 
% the vortex liquid state $\rightarrow$ 
% superconducting state strongly suppressed 
% by a tilt of the field $\rightarrow$ 
% superconducting state sensitive to impurity. 
%
% We should pay attention to the possibility of such three different 
% superconducting states in discussing the pairing competition 
% on the (TMTSF)$_{2}X$, 
% because the pairing competition is very close in the actual materials 
% and the different anion and the additional effect 
% (the magnetic field, the impurity or the pressure) 
% might change the dominant pairing symmetry. 
% Further theoretical and experimental studies about 
% clarifying their difference are required 
% for a more clear understanding of 
% the nature of the superconducting state. 
%@@
% @@ about correspondence 

%%%%%%%%%%%%%%%%%%%%%%%%%%%%%%%%%%%%%%%%%%%%%%%%%%%%%%%%%%%%%%%%%%%%%%%%%%%%%%
%\begin{acknowledgments}
 We acknowledge S. Yonezawa for valuable discussions. 
 We acknowledge Grants-in-Aid for Scientific Research from the
 Ministry of Education, Culture, Sports, Science and Technology of
 Japan, and from the Japan Society for the Promotion of Science.
 Part of the calculation has been performed at the 
 facilities of the Supercomputer Center, 
 ISSP, University of Tokyo.
 T. Y. acknowledges support by the Japan Society for the Promotion 
 of Science (JSPS). 
%\end{acknowledgments}

%\clearpage

%%%%%%%%%%%%%%%%%%%%%%%%%%%%%%%%%%%%%%%%%%%%%%%%%%%%%%%%%%%%%%%%%%%%%%%%%%%%%%

%%%%%%%%%%%%%%%%%%%%%%%%%%%%%%%%%%%%%%%%%%%%%%%%%%%%%%%%%%%%%%%%%%%%%%%%%%

%1\\2\\3\\4\\5

%%%%%%%%%%%%%%%%%%%%%%%%%%%%%%%%%%%%%%%%%%%%%%%%%%%%%%%%%%%%%%%%%%%%%%%%%%
%%%%%%%%%%%%%%%%%%%%%%%%%%%%%%%%%%%%%%%%%%%%%%%%%%%%%%%%%%%%%%%%%%%%%%%%%%
%%%%%%%%%%%%%%%%%%%%%%%%%%%%%%%%%%%%%%%%%%%%%%%%%%%%%%%%%%%%%%%%%%%%%%%%%%
\end{document}